\documentclass[preprint,secnumarabic,amssymb, nobibnotes, aps, prd]{revtex4}
\usepackage{amsmath,amssymb,amsfonts}
\usepackage[dvips]{graphicx}

\newcommand{\agev}    {$A$~GeV}               
\newcommand{\gevc}    {GeV$/c$}

\newcommand{\sdyn}    {$\sigma_{\text{dyn}}$}
\newcommand{\ndyn}    {$\nu_{\text{dyn}}$}
\newcommand{\nw}      {$N_{\text{W}}$}
\newcommand{\anw}     {$\langle N_{\text{W}} \rangle$}
\newcommand{\dEdx}    {d$E$/d$x$}
\newcommand{\pt}      {$p_{\text{T}}$}

\usepackage{type1cm}
\usepackage{eso-pic}
\usepackage{color}
\makeatletter
\AddToShipoutPicture{
\setlength{\@tempdimb}{0.5\paperwidth}
\setlength{\@tempdimc}{0.5\paperheight}
\setlength{\unitlength}{1pt}
\put(\strip@pt\@tempdimb,\strip@pt\@tempdimc){
}
}
\makeatother

\begin{document}

\title{System size dependence of particle-ratio fluctuations
in Pb+Pb collisions at 158\agev}

\author{\noindent
T.~Anticic$^{22}$, B.~Baatar$^{8}$, D.~Barna$^{4}$, J.~Bartke$^{6}$, 
H.~Beck$^{9}$, L.~Betev$^{10}$, H.~Bia{\l}\-kowska$^{19}$, C.~Blume$^{9}$, 
M.~Bogusz$^{21}$, B.~Boimska$^{19}$, J.~Book$^{9}$, M.~Botje$^{1}$,
P.~Bun\v{c}i\'{c}$^{10}$,
T.~Cetner$^{21}$, P.~Christakoglou$^{1}$,
P.~Chung$^{18}$, O.~Chv\'{a}la$^{14}$, J.G.~Cramer$^{15}$, V.~Eckardt$^{13}$,
Z.~Fodor$^{4}$, P.~Foka$^{7}$, V.~Friese$^{7}$,
M.~Ga\'zdzicki$^{9,11}$, K.~Grebieszkow$^{21}$, C.~H\"{o}hne$^{7}$,
K.~Kadija$^{22}$, A.~Karev$^{10}$, V.I.~Kolesnikov$^{8}$, 
T.~Kollegger$^{9}$, M.~Kowalski$^{6}$, D.~Kresan$^{7}$,
A.~L\'{a}szl\'{o}$^{4}$, R.~Lacey$^{18}$, M.~van~Leeuwen$^{1}$,
M.~Ma\'{c}kowiak-Paw{\l}owska$^{21}$, M.~Makariev$^{17}$, A.I.~Malakhov$^{8}$,
M.~Mateev$^{16}$, G.L.~Melkumov$^{8}$, M.~Mitrovski$^{9}$, St.~Mr\'owczy\'nski$^{11}$, 
V.~Nicolic$^{22}$, G.~P\'{a}lla$^{4}$, A.D.~Panagiotou$^{2}$, W.~Peryt$^{21}$, 
J.~Pluta$^{21}$, D.~Prindle$^{15}$,
F.~P\"{u}hlhofer$^{12}$, R.~Renfordt$^{9}$, C.~Roland$^{5}$, G.~Roland$^{5}$,
M. Rybczy\'nski$^{11}$, A.~Rybicki$^{6}$, A.~Sandoval$^{7}$, 
N.~Schmitz$^{13}$, T.~Schuster$^{9}$, P.~Seyboth$^{13}$, F.~Sikl\'{e}r$^{4}$, 
E.~Skrzypczak$^{20}$, M.~S{\l}odkowski$^{21}$, G.~Stefanek$^{11}$, R.~Stock$^{9}$, 
H.~Str\"{o}bele$^{9}$, T.~Susa$^{22}$, M.~Szuba$^{21}$, 
M.~Utvi\'{c}$^{9}$, D.~Varga$^{3}$, M.~Vassiliou$^{2}$,
G.I.~Veres$^{4}$, G.~Vesztergombi$^{4}$, D.~Vrani\'{c}$^{7}$,
Z.~W{\l}odarczyk$^{11}$, A.~Wojtaszek-Szwar\'{c}$^{11}$
}

\affiliation{
\noindent
$^{1}$ NIKHEF, Amsterdam, Netherlands. \\
$^{2}$ Department of Physics, University of Athens, Athens, Greece.\\
$^{3}$ E\"otv\"os Lor\'ant University, Budapest, Hungary \\
$^{4}$ Wigner Research Center for Physics, Hungarian Academy of Sciences, Budapest, Hungary.\\
$^{5}$ MIT, Cambridge, Massachusetts, USA.\\
$^{6}$ H.~Niewodnicza\'nski Institute of Nuclear Physics, Polish Academy of Sciences, Cracow, Poland.\\
$^{7}$ GSI Helmholtzzentrum f\"{u}r Schwerionenforschung GmbH, Darmstadt, Germany.\\
$^{8}$ Joint Institute for Nuclear Research, Dubna, Russia.\\
$^{9}$ Fachbereich Physik der Universit\"{a}t, Frankfurt, Germany.\\
$^{10}$ CERN, Geneva, Switzerland.\\
$^{11}$ Institute of Physics, Jan Kochanowski University, Kielce, Poland.\\
$^{12}$ Fachbereich Physik der Universit\"{a}t, Marburg, Germany.\\
$^{13}$ Max-Planck-Institut f\"{u}r Physik, Munich, Germany.\\
$^{14}$ Institute of Particle and Nuclear Physics, Charles University, Prague, Czech Republic.\\
$^{15}$ Nuclear Physics Laboratory, University of Washington, Seattle, Washington, USA.\\
$^{16}$ Atomic Physics Department, Sofia University St. Kliment Ohridski, Sofia, Bulgaria.\\
$^{17}$ Institute for Nuclear Research and Nuclear Energy, BAS, Sofia, Bulgaria.\\
$^{18}$ Department of Chemistry, Stony Brook University (SUNYSB), Stony Brook, New York, USA.\\
$^{19}$ Institute for Nuclear Studies, Warsaw, Poland.\\
$^{20}$ Institute for Experimental Physics, University of Warsaw, Warsaw, Poland.\\
$^{21}$ Faculty of Physics, Warsaw University of Technology, Warsaw, Poland.\\
$^{22}$ Rudjer Boskovic Institute, Zagreb, Croatia.\\
}

\date{\today }

\begin{abstract}
New measurements by the NA49 experiment of the centrality dependence of
event-by-event fluctuations of the particle yield ratios  
(K$^{+}$+K$^{-}$)/($\pi^{+}$+$\pi^{-}$),
(p+$\bar{\text p}$)/($\pi^{+}$+$\pi^{-}$),  
and (K$^{+}$+K$^{-}$)/(p+$\bar{\text p}$)
are presented for Pb+Pb collisions at 158\agev.
The absolute values of the dynamical fluctuations of these ratios, 
quantified by the measure \sdyn, increase by about a factor of two 
from central to semi-peripheral collisions. 
Multiplicity scaling scenarios are tested and found to apply for
both the centrality and the previously published energy dependence 
of the (K$^{+}$+K$^{-}$)/($\pi^{+}$+$\pi^{-}$) and (p+$\bar{\text p}$)/($\pi^{+}$+$\pi^{-}$)
ratio fluctuations. A description of the centrality and energy dependence of
(K$^{+}$+K$^{-}$)/(p+$\bar{\text p}$) ratio fluctuations by a common scaling prescription
is not possible since there is a sign change in the energy dependence.
\end{abstract}

\maketitle


\section{Introduction}
 
The search for structures in the QCD phase diagram, like the first order phase
transition line from hadronic to partonic degrees of freedom or the critical
endpoint, has become one of the main activities in current and future
high-energy heavy-ion experiments \cite{Shuryak:2001ik,na61prop,starprop}. The experimental
signatures for these
structures are the subject of ongoing discussions. Lattice QCD
calculations show that in the co-existence region of hadronic and partonic
degrees of freedom and in the vicinity of the
critical endpoint event-by-event fluctuations of, for example, the 
strangeness-to-entropy ratio increase significantly 
\cite{Karsch:CPOD2007,Cheng:2009,Schmidt:CPOD2009,Machado:2010wi}. Thus,
a measurement of the energy dependence of a quantity sensitive to this
ratio and an observation of a non-monotonic behavior may provide
an indication of the location of the critical endpoint.

The NA49 experiment at the CERN Super Proton Synchrotron (SPS) analyzed the energy dependence of the ratio of inclusive
K$^{+}$ and $\pi^{+}$ yields in central Pb + Pb collisions and observed a peak
structure at beam energies around 30 - 40\agev~\cite{na49:deconf}. This motivated the analysis of
event-by-event fluctuations of the (K$^{+}$+K$^{-}$)/($\pi^{+}$+$\pi^{-}$)
(denoted K/$\pi$)~\cite{bib:na49:ebefluct},  
(p+$\bar{\text p}$)/($\pi^{+}$+$\pi^{-}$) (denoted p/$\pi$)~\cite{bib:na49:ebefluct} 
and (K$^{+}$+K$^{-}$)/(p+$\bar{\text p}$) (denoted K/p) ratios~\cite{bib:na49:tim} as function of 
the center-of-mass energy by means of the observable \sdyn (see see Eqs.~\ref{eq:sigma},\ref{eq:sdyn} in section 3.4),
which measures the dynamical contribution to the fluctuations of the event-by-event particle ratios. 
The K/$\pi$ ratio fluctuations show a continuous increase towards lower collision
energies, which is not reproduced by the UrQMD model~\cite{urqmd}, but obtained
qualitatively by HSD model calculations~\cite{hsd}. The p/$\pi$ ratio fluctuations as
a function of the center-of-mass energy show negative values which
indicate strong correlations. This observation is 
well reproduced by UrQMD model calculations and can be
interpreted as the result of the production of nucleon resonances and
their decays into pions and protons. 
The K/p ratio fluctuations exhibit a change of sign at $\approx~$30\agev~beam energy which is
not well understood \cite{bib:na49:tim}. In view of the complex energy
dependence of the fluctuations of the three particle ratios an
additional study of their collision centrality dependence at the top SPS energy
may help to clarify the interpretation. In particular, such an investigation may help to distinguish the
contributions of the changing multiplicities and the genuine energy and collision volume 
dependence of the underlying correlations~\cite{Koch:2009dg}.

The STAR collaboration at the Relativistic Heavy Ion Collider (RHIC)
also published results on particle-ratio fluctuations~\cite{Abelev:kpifluct} employing the
observable \ndyn~(Eq.~\ref{eq:sdyn:nudyn} in section 3.4). First results from a recent low energy scan in Au+Au collisions 
at $\sqrt{s_{NN}}=7.7$~GeV were presented at conferences~\cite{Star:qm2011} and 
show a different trend for the energy dependence of K/$\pi$ and K/p fluctuations when compared using the 
equivalence relation between \ndyn~and \sdyn~(see Eq.~\ref{eq:sdyn:nudyn} in section 3.4). 
However, acceptance in both rapidity $y$ and transverse momentum \pt~as
well as the selection procedure of collision centrality differ.

In this paper we address the dependence of event-by-event fluctuations of
particle yield ratios on the centrality of Pb+Pb collisions in a fixed
acceptance and at a beam energy of 158\agev~\cite{bib:dkresan}. In section 2 we describe the
experimental equipment, in section 3 the analysis procedures. Section 4 presents
the experimental results and compares to various
proposed multiplicity scaling schemes. A summary section 5 closes
the paper.

\section{The NA49 experiment}

NA49 is a fixed target experiment \cite{Afanasev:1999iu} at the CERN SPS.
The trajectories of
charged particles are reconstructed in four large volume Time Projection
Chambers (TPCs). Two of them (VTPCs) are placed inside of two superconducting dipole
magnets for momentum determination. Two main TPCs (MTPCs) are located
downstream of the magnets on both sides of the beam. The performance of the
MTPCs is tuned for high precision measurements of the specific energy loss
\dEdx, which is the basis for particle identification employed in this
analysis (see section \ref{sec:dedx} and \cite{Wenig:1998vv}). Except for the
trigger and beam intensity the experimental conditions in this analysis are the
same as described in \cite{bib:na49:ebefluct}. The Pb beam had a typical
intensity of 10$^{4}$~ions/s. The minimum bias trigger was derived from a
He-Cerenkov counter placed behind the target.  Only interactions which reduced
the beam charge and thus the signal seen by this detector by at least 10\%, were
accepted. The trigger cross section thus defined is 5.7~b out of a total inelastic
cross section of 7.15~b. 
The resulting ensemble of 174 K events was divided into centrality classes 
according to the energy measured in the Veto Calorimeter (VCAL)
located 26~m downstream from the target.


\section{Data analysis}

\subsection{Event, track selection and acceptance}

In order to reject backgound interactions, a valid fit of an event
vertex was required and a cut around the known target position was applied.
The contamination by background events remaining after cuts on vertex position and
quality amounts to less than 5\% for the most peripheral collisions and is
negligible for near-central collisions.

The useful acceptance for pions, kaons, and protons is constrained by the needs of
particle identification. The separation power is highest for particles with large track
lengths in the MTPCs which limits the analysis to the forward hemisphere in
the center-of-mass frame. The coverage in the azimuthal angle $\phi$
is a function of center-of-mass rapidity $y$ and transverse momentum \pt.
The loose and tight sets of track cuts used
in the present analysis are given in Table~\ref{tab:cuts}. These are identical to
those employed previously in NA49 analyses of fluctuations~\cite{bib:na49:ebefluct}.
The acceptance is not only determined by the track selection cuts. In addition, only phase
space bins are used for which the inclusive \dEdx~distributions have more than
3000 entries.

\begin{table}[ht]
\caption{Loose and tight set of track cuts used in the analysis.}
\centering
\begin{tabular}{l c c}
\hline
\hline
Cut description & \multicolumn{2}{c}{Cut} \\
         & Loose & Tight \\
\hline
(\dEdx)/(\dEdx)$_{\text{MIP}}$ & $\leq$~1.8 & $\leq$~1.8 \\
Number of points in MTPC & $>$~30 & $>$~30 \\
Number of points in VTPC1 & -- & $>$~10 \\
Number of points in VTPC2 & -- & $>$~10 \\
Fraction of potential points found in MTPC & $\geq$~50\% & $\geq$~50\% \\
Number of entries required in phase space & $>$3000 & $>$3000 \\
 bin for fit of inclusive \dEdx~distribution & & \\
Cut in proton rapidity for \pt$\leq~0.2$~\gevc~& $y < y_{\text{beam}}-1$ & $y < y_{\text{beam}}-1$ \\
Track fitted to primary vertex & -- & yes \\
impact parameter $x$-projection & -- & $<$~4~cm \\
impact parameter $y$-projection & -- & $<$~0.5~cm \\
\hline
\hline
\end{tabular}
\label{tab:cuts}
\end{table}

\begin{figure}[ht]
\centering
\includegraphics[width=0.35\textwidth]{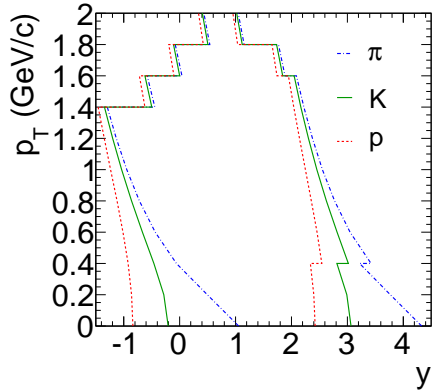}
\caption{(Color online) Acceptance in center-of-mass rapidity $y$ and 
transverse momentum \pt~for the most central Pb+Pb collisions at 158\agev.}
\label{fig:acc}
\end{figure}

{\bf The acceptance after all selection cuts is shown in Fig.~\ref{fig:acc} for
central collisions. The range in \pt~varies slightly depending on the
number of events in the centrality bin.}

\subsection{Collision centrality determination}

\begin{figure}[ht]
\centering
\includegraphics[width=0.35\textwidth]{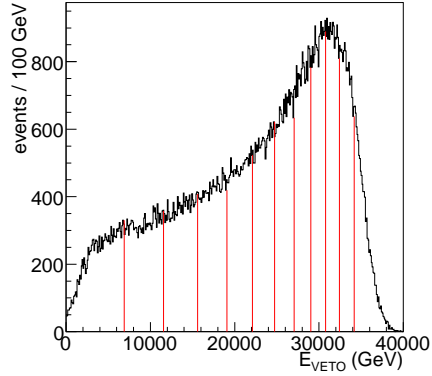}
\caption{(Color online) Distribution of the total energy $E_{\text{VETO}}$ of the projectile spectators
deposited in the VCAL of NA49 in Pb + Pb collisions at 158\agev~beam energy. An event vertex cut
(see text) was applied to remove background triggers.
Vertical lines separate bins of 5\% of the total inelastic cross section.}
\label{fig:eveto}
\end{figure}

The determination of the centrality of the collisions 
is based on the energy of forward going projectile
spectators as measured in the VCAL.
The distribution of the VCAL energy $E_{\text{VETO}}$ together with the
division into 5\% bins of the total inelastic cross section 
is shown in Fig. \ref{fig:eveto}.

The energy resolution of the VCAL measurement is dominated by two effects: the
intrinsic energy resolution as given by the longitudinal sampling structure and by the
non-uniformity of light collection efficiency.
The overall resolution of the calorimeter was shown to follow~\cite{Afanasev:1999iu}:

\begin{equation}
\frac{\sigma_{E}}{E} \approx \frac{2}{\sqrt{E}} \text{~~,} 
\label{eq:vcal:se}
\end{equation} 
with $E$ in units of GeV.

\begin{figure}[ht]
\centering
\includegraphics[width=0.35\textwidth]{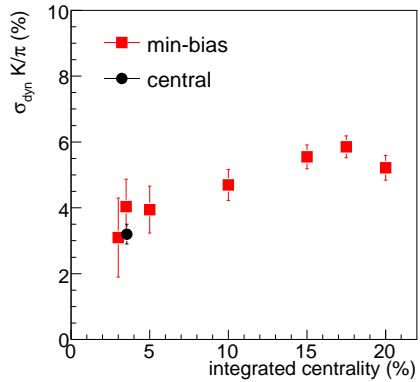}
\caption{(Color online) Dependence of the \sdyn~measure of K/$\pi$ ratio fluctuations 
for the most central collisions on the width of the centrality bin in Pb+Pb collisions at 158\agev.}
\label{fig:cbins}
\end{figure}

The choice of 5\% centrality bin size is motivated by the energy resolution of
VCAL, the requirement to keep the reaction volume fluctuations at a minimum and the
necessity to have sufficient statistics in each centrality bin.
Volume fluctuations are relevant for ratios involving kaons, since their
multiplicity does not strictly scale with the number of wounded nucleons \nw, or
equivalently the reaction volume, in contrast to the multiplicity of pions
and protons~\cite{na49:centkpi}. The influence of volume fluctuations on all particle-ratio
fluctuations was studied by varying the centrality bin widths in the
range from 3\% - 20\%. The results shown in Fig.~\ref{fig:cbins} for the example of
K/$\pi$ ratio fluctuations in the most central collisions led us to choose 5\% wide centrality bins,
the smallest bin size that leaves sufficient satistics. 
For each bin the corresponding average number of wounded nucleons \anw~was obtained
from the Glauber model approach using a simulation with the VENUS event 
generator~\cite{venus,laszlo}.

\subsection{Particle identification by \dEdx}
\label{sec:dedx}

The event-by-event measurement of particle ratios ideally implies 
track-by-track identification of the
different particle types. The NA49 experiment provides energy loss measurements
along the particle trajectories in the MTPCs with a resolution of approximately
4~\% in the relativistic rise region for particle momenta $p$ above 3~\gevc. Since
the separation of the \dEdx~signals of pions, kaons, and protons at a given
momentum is of the same order, track-by-track identification of particle types
is not possible. Instead, we employ a statistical method, namely the Maximum  
Likelihood Method (MLM) to extract particle ratios from event-wise
\dEdx~distributions of negatively and of positively charged particles.

In a first step energy loss distributions of all accepted tracks in the
event ensemble were constructed in bins of $p$, $p_{T}$
and $\phi$. The binning details are shown in Table~\ref{tab:bins}.

\begin{table}[ht]
\caption{Binning in phase space used for fitting the inclusive
\dEdx~distributions. Due to overlap of the distributions for
different particle species around momenta of 3~\gevc~the first 3 bins in
total momentum were not used.}
\centering
\begin{tabular}{c c c c}
\hline
\hline
Variable & Range & $N_{\text{bins}}$ & Bin size \\
\hline
$p$ & 1-120~\gevc & 20 & logarithmic \\
\pt & 0-2~\gevc & 10 & 0.2~\gevc \\
$\phi$ & 0-2$\pi$ & 8 & 0.25$\cdot\pi$ \\
charge $q$ & 1,-1 & 2 & - \\
\hline
\hline
\end{tabular}
\label{tab:bins}
\end{table}

The resulting inclusive specific energy loss distributions 
in each phase space bin were fitted with four Gaussian functions all having
the same width for electrons, pions, kaons, and protons (and their
antiparticles). The values of the nine fit parameters
(eight positions and one width) define Probability Density Functions (PDF)
for each phase space bin and were stored in a look-up table for later use 
in the event-by-event fits.

Using the PDFs one can calculate for each particle the
four probabilities $f_{\alpha}$ to be an electron (positron), a pion, a kaon or
a proton (anti-proton). The sum of these probabilities weighted
with coefficients $\theta_{\alpha}$ become the factors in the likelihood
function which depends on the coefficients
$\theta_{\alpha}$:

\begin{equation}
L(\{\theta_{\alpha}\}) = \prod_{i=1}^{N}\sum_{\alpha}\theta_{\alpha}f_{\alpha}(q^{i},p^{i},p_{t}^{i},\phi^{i},(dE/dx)^{i}) \text{~~,}
\label{eq:likelihood}
\end{equation}
where the index i runs over the $N$ particles of the event. 
The coefficients $\theta_{\alpha}$ are the relative yield fractions of each
particle type in the event. The sum of the weights is constrained to unity:

\begin{equation}
\sum_{\alpha}\theta_{\alpha} = 1 \text{~~.}
\label{eq:theta}
\end{equation}

By maximizing the likelihood function with respect to the relative yield
fractions one obtains the best estimate of the different particle
multiplicities in a given event.
More details about the employed MLM can be found in \cite{bib:croland}.

\subsection{Extraction of dynamical fluctuations}

The fluctuations of particle ratios in the event ensemble are defined as the ratio of the
root of the variance $\sqrt{\text{Var}(A/B)}$ of the distribution of the event-wise particle yield ratio $A/B$  to
the mean $\langle A/B \rangle$ of the same distribution:

\begin{equation}
\sigma = \frac{\sqrt{\text{Var}(A/B)}}{\langle A/B \rangle} \text{~~.}
\label{eq:sigma}
\end{equation}

\begin{figure}[ht]
\centering
\includegraphics[width=0.35\textwidth]{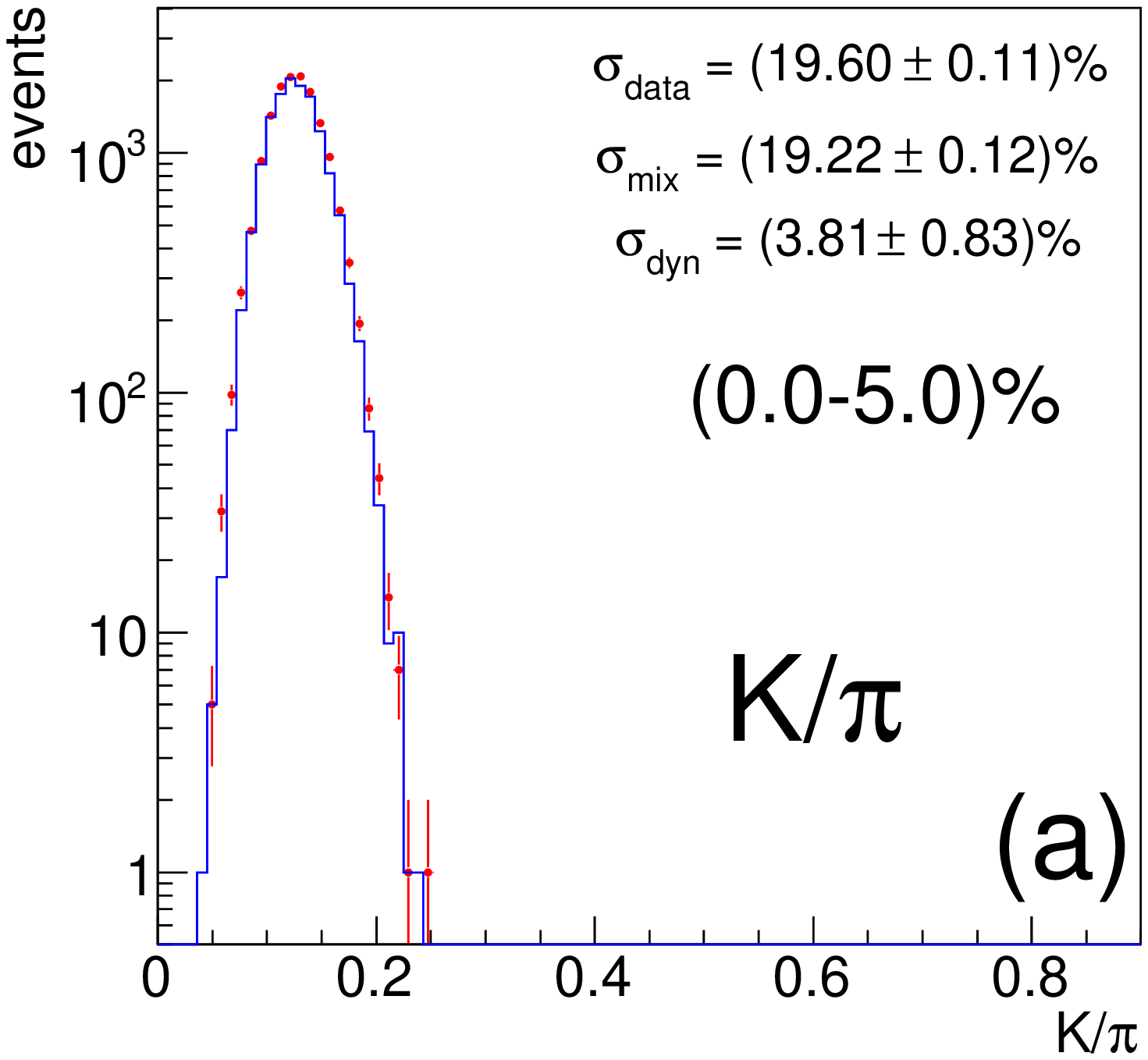}
\includegraphics[width=0.35\textwidth]{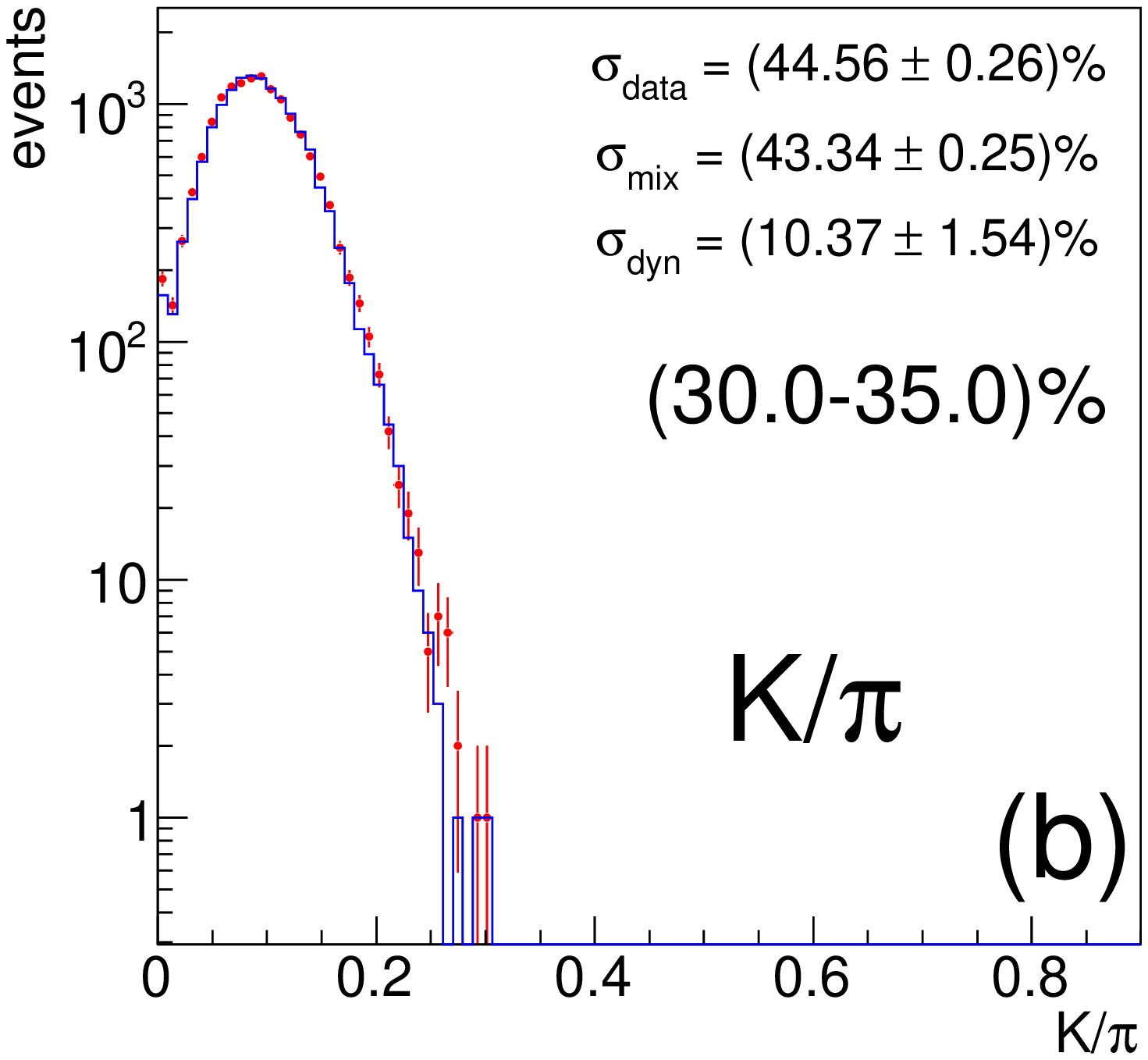} \\
\includegraphics[width=0.35\textwidth]{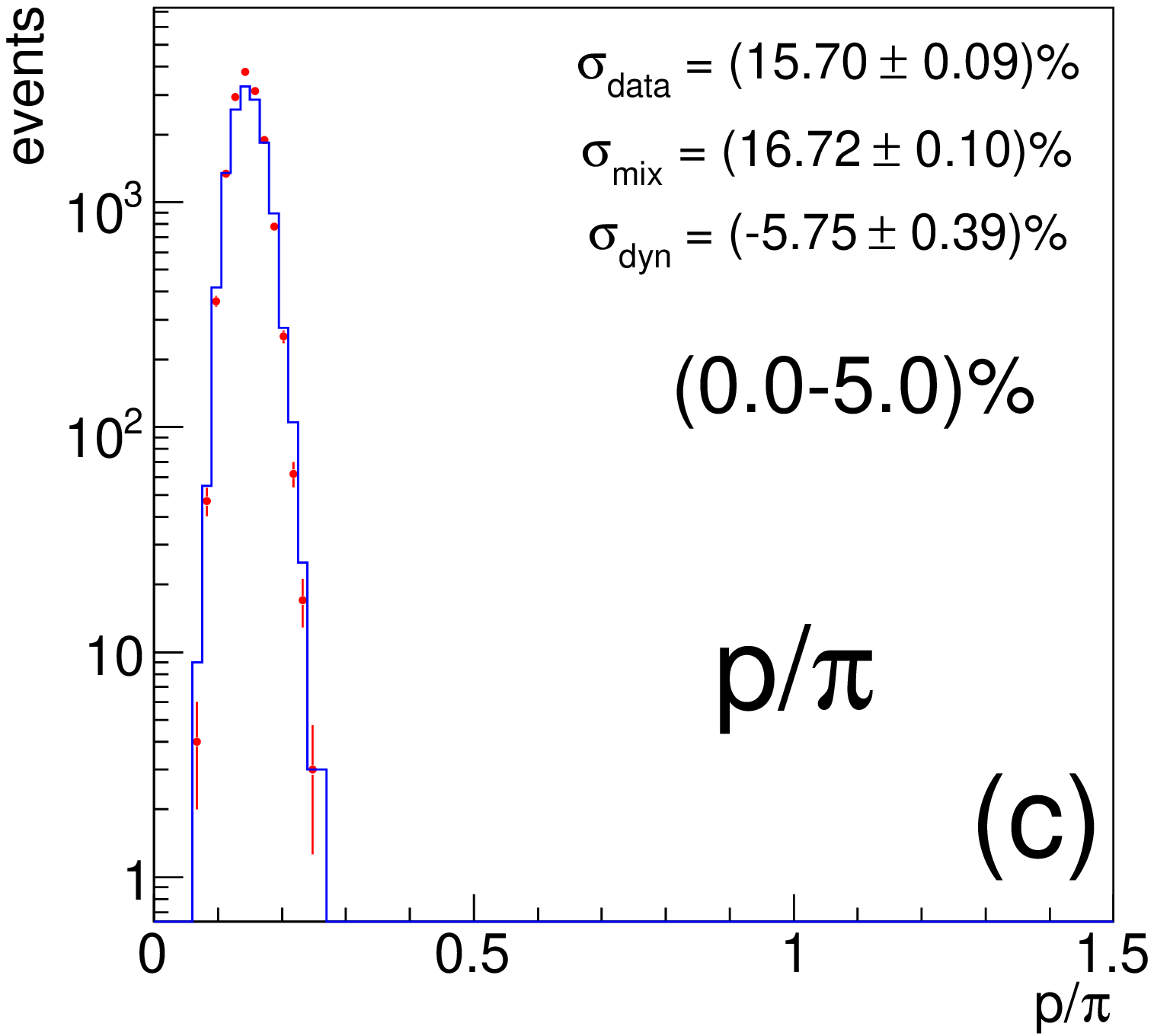}
\includegraphics[width=0.35\textwidth]{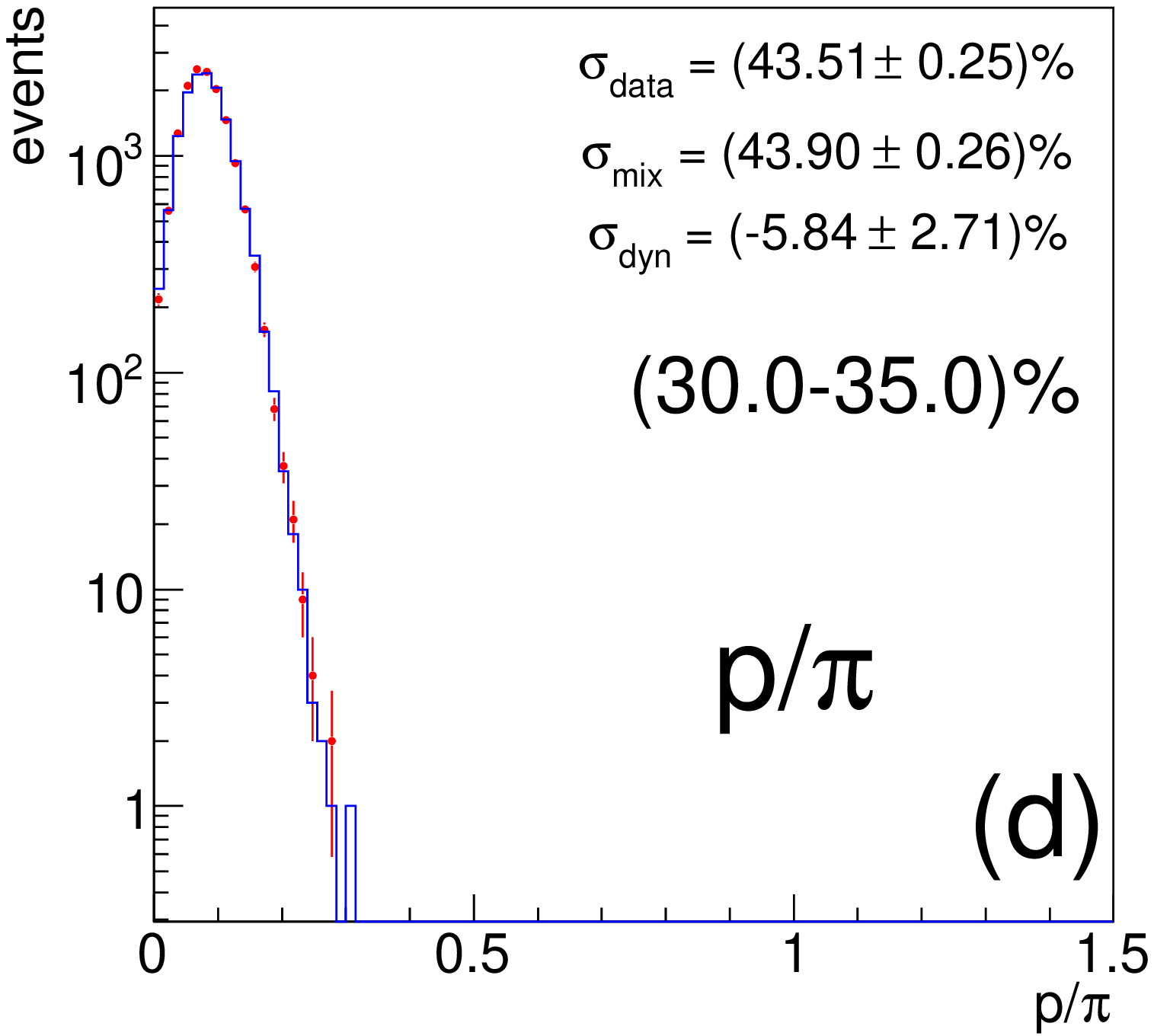} \\
\includegraphics[width=0.35\textwidth]{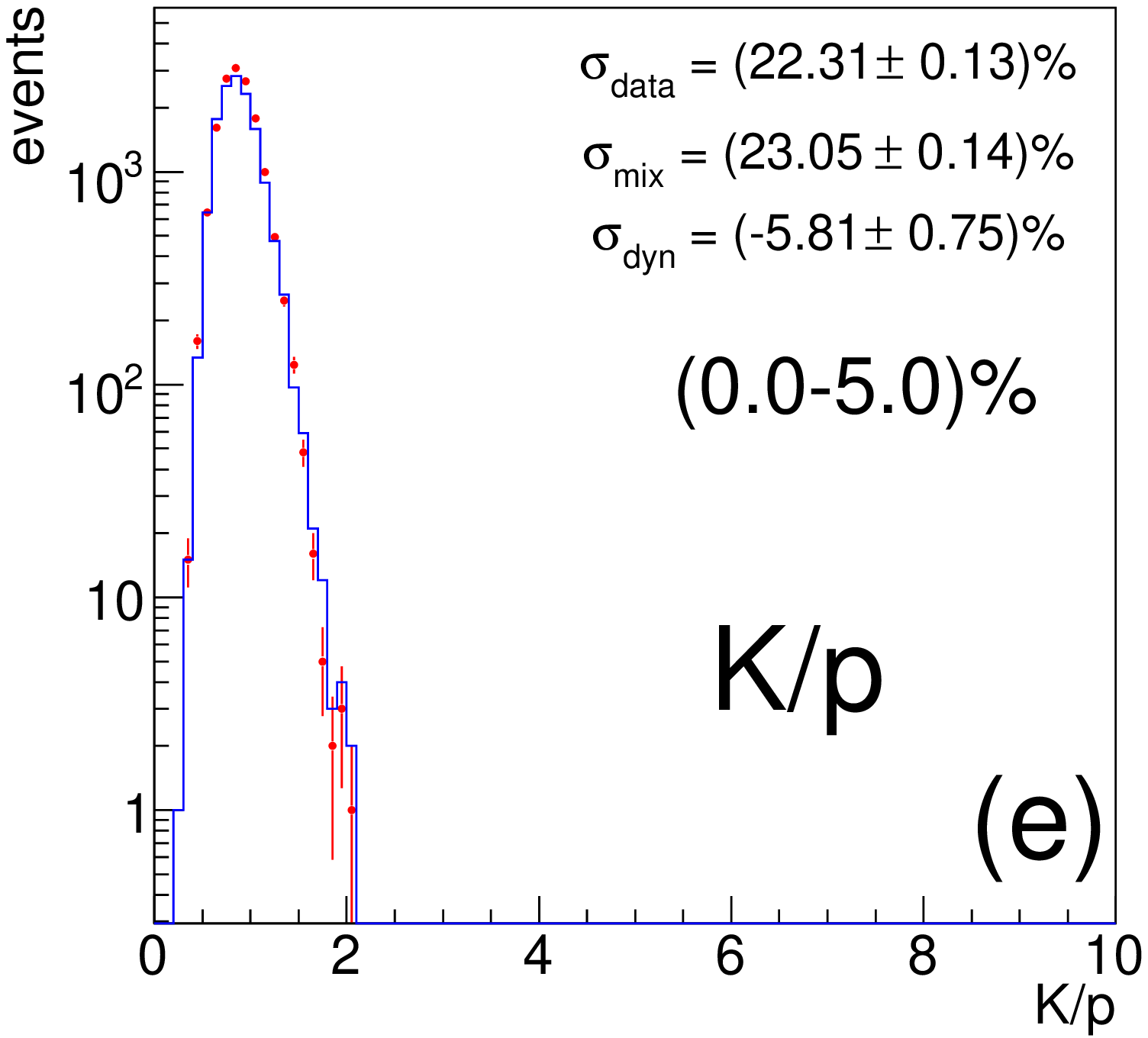}
\includegraphics[width=0.35\textwidth]{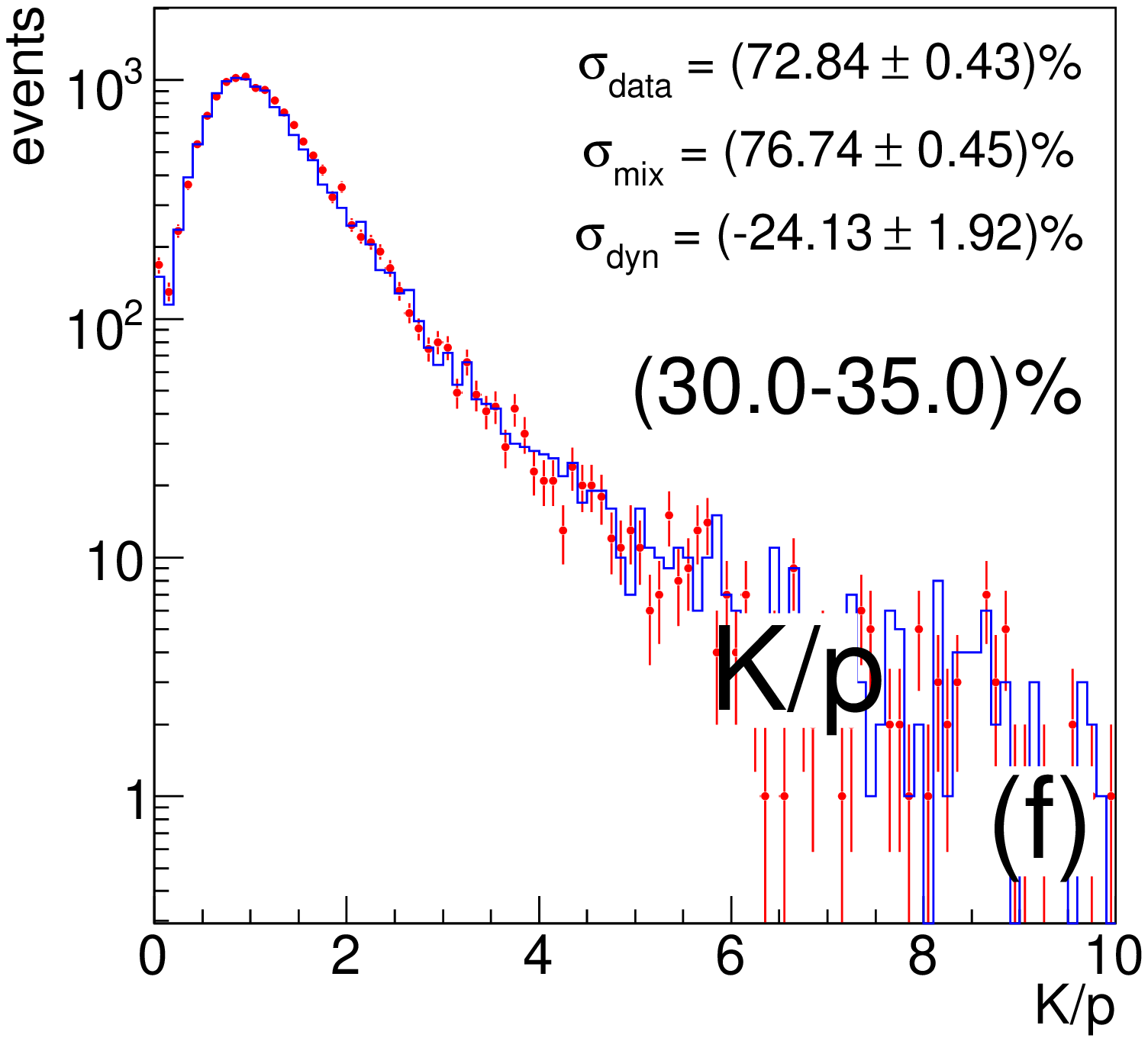}
\caption{(Color online) Event-by-event particle-ratio distributions for 
central (left) and semi-peripheral (right) in Pb+Pb collisions at 158\agev. The data points show real event and the histogram
mixed event distributions. Loose track cuts (see~Table~\ref{tab:cuts}) were applied.}
\label{fig:diff:bins}
\end{figure}

Defined in this way $\sigma_{\text{data}}$ will contain contributions from the finite number
statistics, detector resolution, nonperfect particle identification 
and the genuine dynamical fluctuations. The first three contributions are considered as
background. Since their contributions dominate the ratio fluctuation signal,
their magnitudes have to be determined quantitatively. For an estimate of the
statistical fluctuations and the detector resolution effects the event mixing  
method was applied. A new ensemble
of artificial events was generated which contain particles from
different real events, selected randomly such that in each artificial event
no pair of particles originates from the same data event. In addition the
multiplicity distribution of the mixed events was constructed to be the same as the corresponding
distribution of the real events. By this token dynamical fluctuations,
which may be present in data, are absent in the sample of mixed events. The
measure $\sigma_{\text{mix}}$, evaluated according to Eq.~\ref{eq:sigma} for the mixed events,
contains thus only the background fluctuations. Examples of distributions of the 
event-wise particle ratio for real and mixed events are shown in Fig.~\ref{fig:diff:bins}
for central and semi-peripheral Pb+Pb collisions.

We now define dynamical fluctuations (\sdyn) as the geometrical
difference between the fluctuations measured in real and mixed events:
\begin{equation}
\sigma_{\text{dyn}} = \text{sign}(\sigma_{\text{data}}^2-\sigma_{\text{mix}}^2) \sqrt{\big|\sigma_{\text{data}}^{2}-\sigma_{\text{mix}}^{2}\big|} \text{~~.}
\label{eq:sdyn}
\end{equation}
Alternatively, the observable \ndyn~\cite{Pruneau:fluct}, defined as:
\begin{equation}
\begin{split}
 & \nu_{\text{dyn}} =  \nu - \nu_{\text{stat}} \\
 & \nu = \frac{\text{Var}(A)}{\langle A \rangle^{2}}+\frac{\text{Var}(B)}{\langle B \rangle^{2}}-
2\frac{\text{Cov}(A,B)}{\langle A \rangle \langle B \rangle}
\end{split}
\label{eq:nudyn}
\end{equation}
has been used to measure dynamical fluctuations of the particle ratio A/B. 
Here $\nu_{\text{stat}} = 1/\langle A \rangle + 1/\langle B \rangle$ is the contribution
from finite number statistics. Assuming that detector effects cancel in 
\sdyn~it was shown that \sdyn~is related~\cite{Abelev:kpifluct,Koch:2009dg} to 
the fluctuation measure \ndyn: 
\begin{equation}
\text{sign}(\sigma_{\text{dyn}}) \sigma_{\text{dyn}}^2 \approx \nu_{\text{dyn}} =
\frac{\text{Var}(A)-\langle A \rangle}{\langle A \rangle^{2}}+
\frac{\text{Var}(B)-\langle B \rangle}{\langle B \rangle^{2}}-
2\frac{\text{Cov}(A,B)}{\langle A \rangle \langle B \rangle} \text{~~.}
\label{eq:sdyn:nudyn}
\end{equation}
For a check of the sytematic uncertainties inherent in the mixed event
background subraction procedure we also determined \ndyn~from our data.
Owing to our non-perfect particle identification we again use mixed events to account for
the background in the evaluation of \ndyn~from the event-by-event fitted particle multiplicities: 
\begin{equation}
\nu_{\text{dyn}} = \nu_{\text{data}} - \nu_{\text{mix}} \text{~~.}
\label{eq:nudyn;na49}
\end{equation}
The resulting values for \ndyn~and $\sigma_{\text{dyn}}^2$~were found to satisfy the
equality of Eq.~\ref{eq:sdyn:nudyn} within the systematic uncertainties estimated for \sdyn.

As can be seen from Eqs.~\ref{eq:sdyn} and \ref{eq:sdyn:nudyn} 
the values of \sdyn~and \ndyn~can be 
positive as well as negative. Assuming Poissonian single particle
distributions, correlations lead to negative values of \sdyn, while
positive values are indicative of anticorrelations between the particles.

\subsection{Systematic error estimation}

In order to study the systematic uncertainties introduced by the track selection
the results from the tight and loose sets of cuts 
(see Table~\ref{tab:cuts}) were compared.
We take the absolute difference between the results of the analysis with the two
extreme conditions as an estimate of the corresponding systematic error.

\begin{figure}[ht]
\centering
\includegraphics[width=0.3\textwidth]{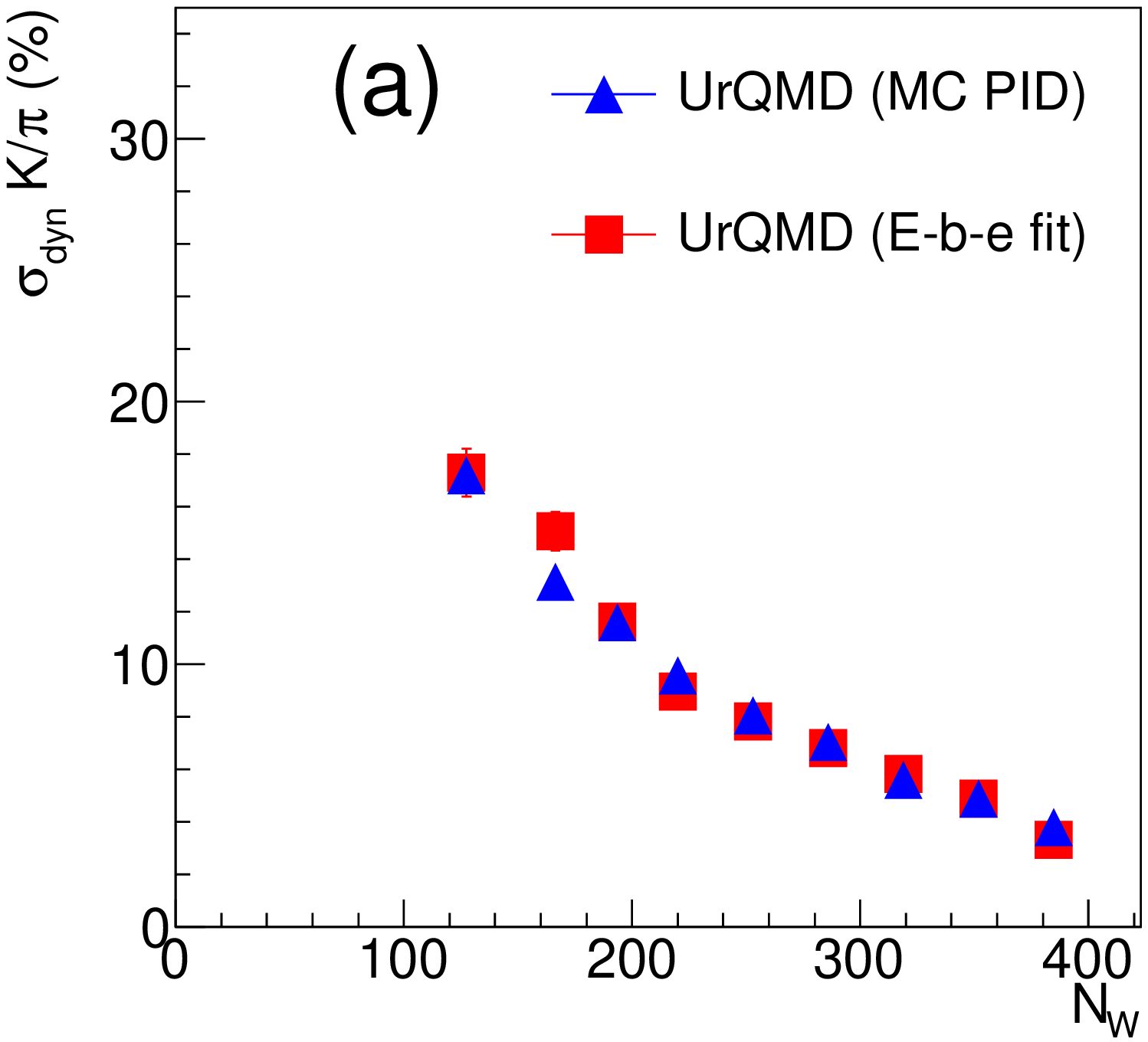} 
\includegraphics[width=0.3\textwidth]{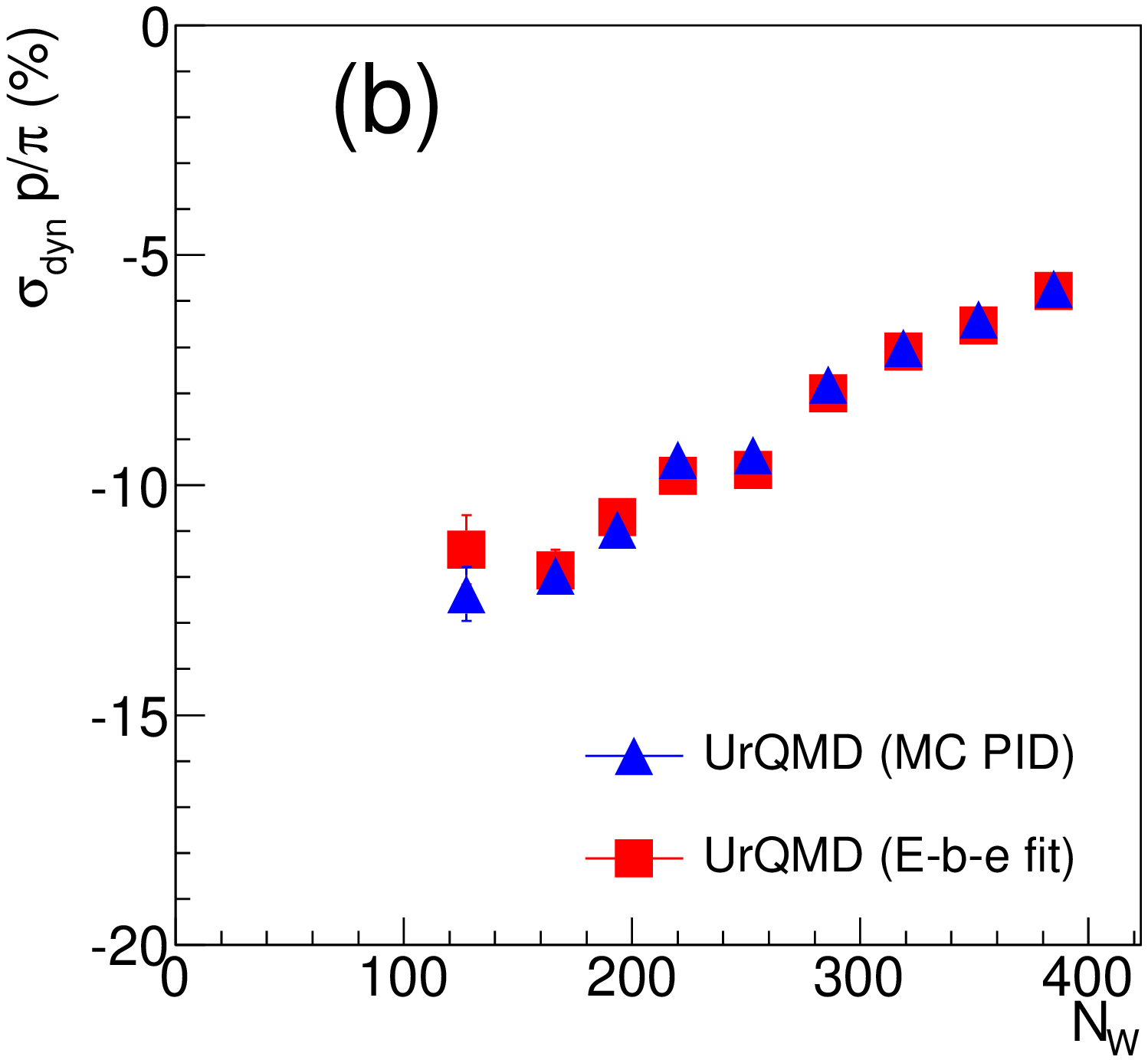} 
\includegraphics[width=0.3\textwidth]{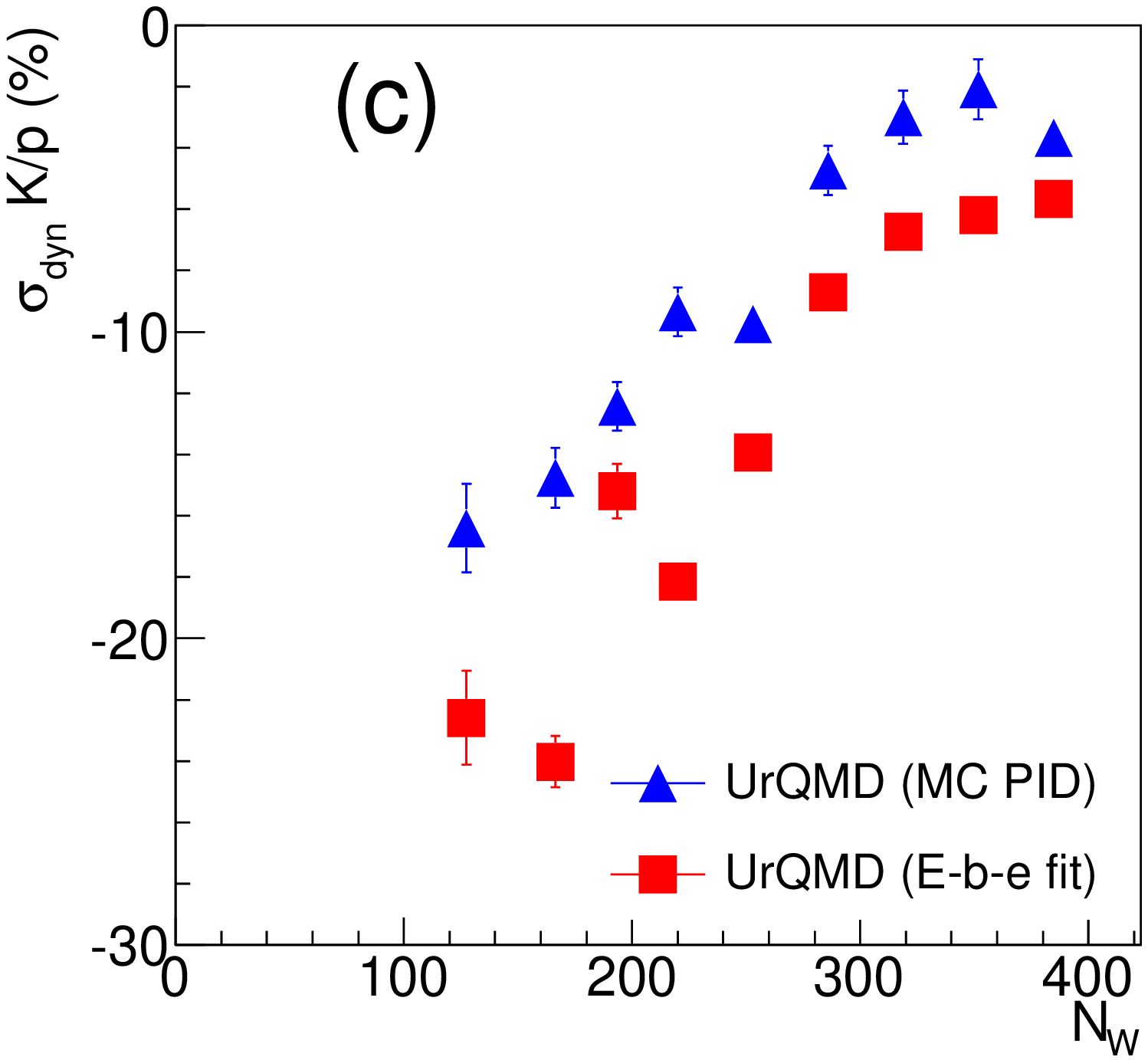}
\caption{(Color online) Centrality dependence of the measure $\sigma_{dyn}$
 for K/$\pi$ (a), p/$\pi$ (b) and K/p (c)
 ratio fluctuations evaluated for events simulated by the UrQMD
 model~\cite{urqmd} using either true identity or the event-by-event
 fit results based on the simulated particle \dEdx.
}
\label{fig:urqmd}
\end{figure}

Other sources of systematic uncertainty for the determination of the particle-ratio fluctuations are the
$dE/dx$~ resolution and the method of event-by-event particle identification.
This systematic effect was studied with the help of simulated events from the UrQMD model~\cite{urqmd}.
In a first step the particles from the generated events were filtered by an 
acceptance table, representing the phase space bins of the real data which
had sufficient statistics for successful fits
of the \dEdx~distribution. Then for each
accepted track a \dEdx~value was randomly generated from a parametrization
of the inclusive \dEdx~distribution for the true particle identity which
depends on particle type and phase space bin. Finally,
the accepted tracks with simulated \dEdx~values were processed by the
same analysis routines as the tracks from real data. In addition to the
determination of particle multiplicities by the MLM the true particle identities
as generated by the Monte Carlo code were stored.
Figure~\ref{fig:urqmd} presents a comparison of the 
values of the dynamical particle-ratio fluctuations as obtained
by using Monte Carlo identity and results from the \dEdx~fit.

The comparison of the results from both identification methods suggests 
the particle identification method based on the event-by-event MLM fit of the 
\dEdx~distributions is valid in the \anw~range above 190. The difference
observed for the K/p ratio was included in the systematic error.


\section{Experimental results and discussion}

\subsection{Centrality dependence of dynamical particle-ratio fluctuations}

\begin{figure}[ht]
\centering
\includegraphics[width=0.3\textwidth]{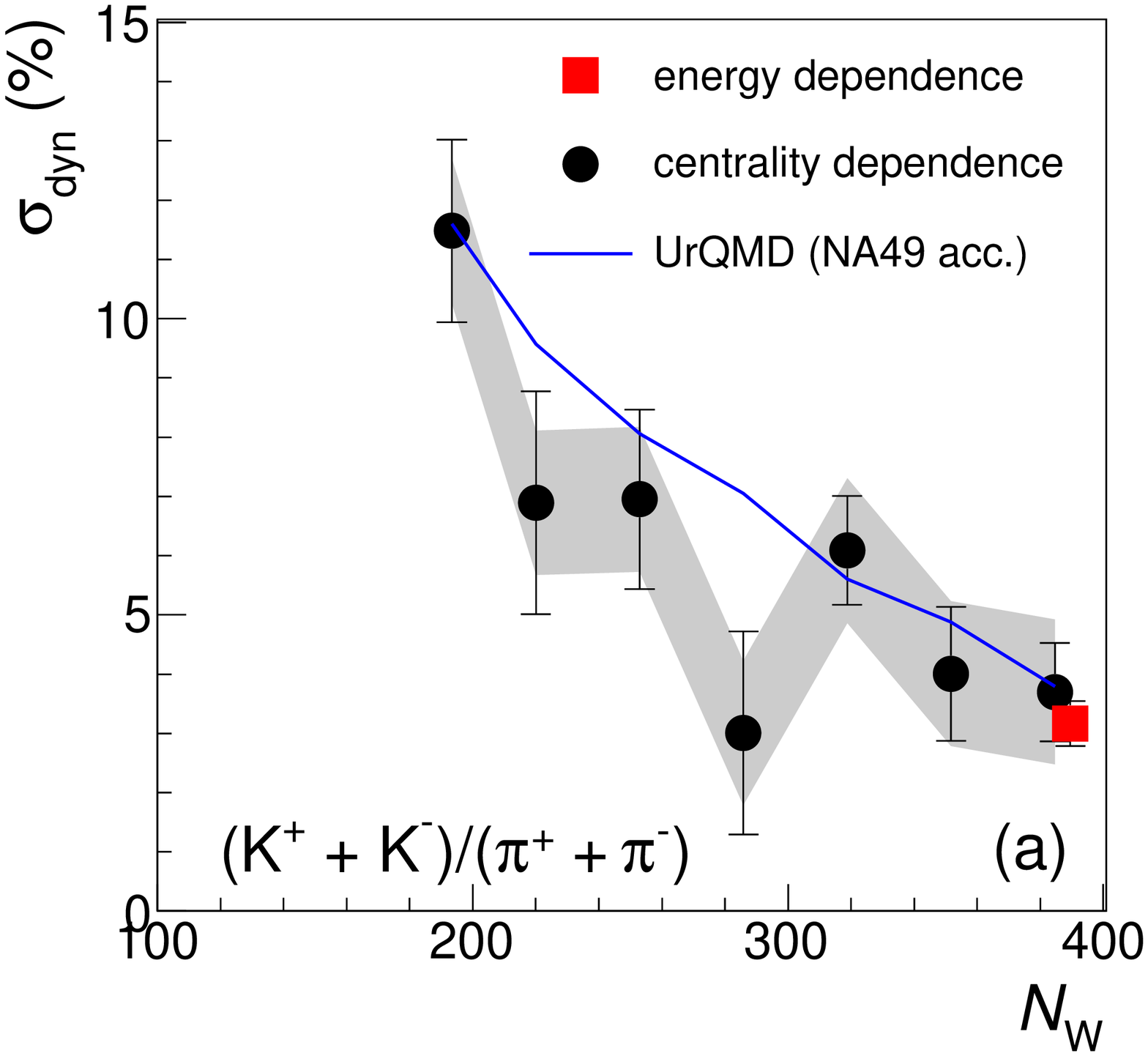}
\includegraphics[width=0.3\textwidth]{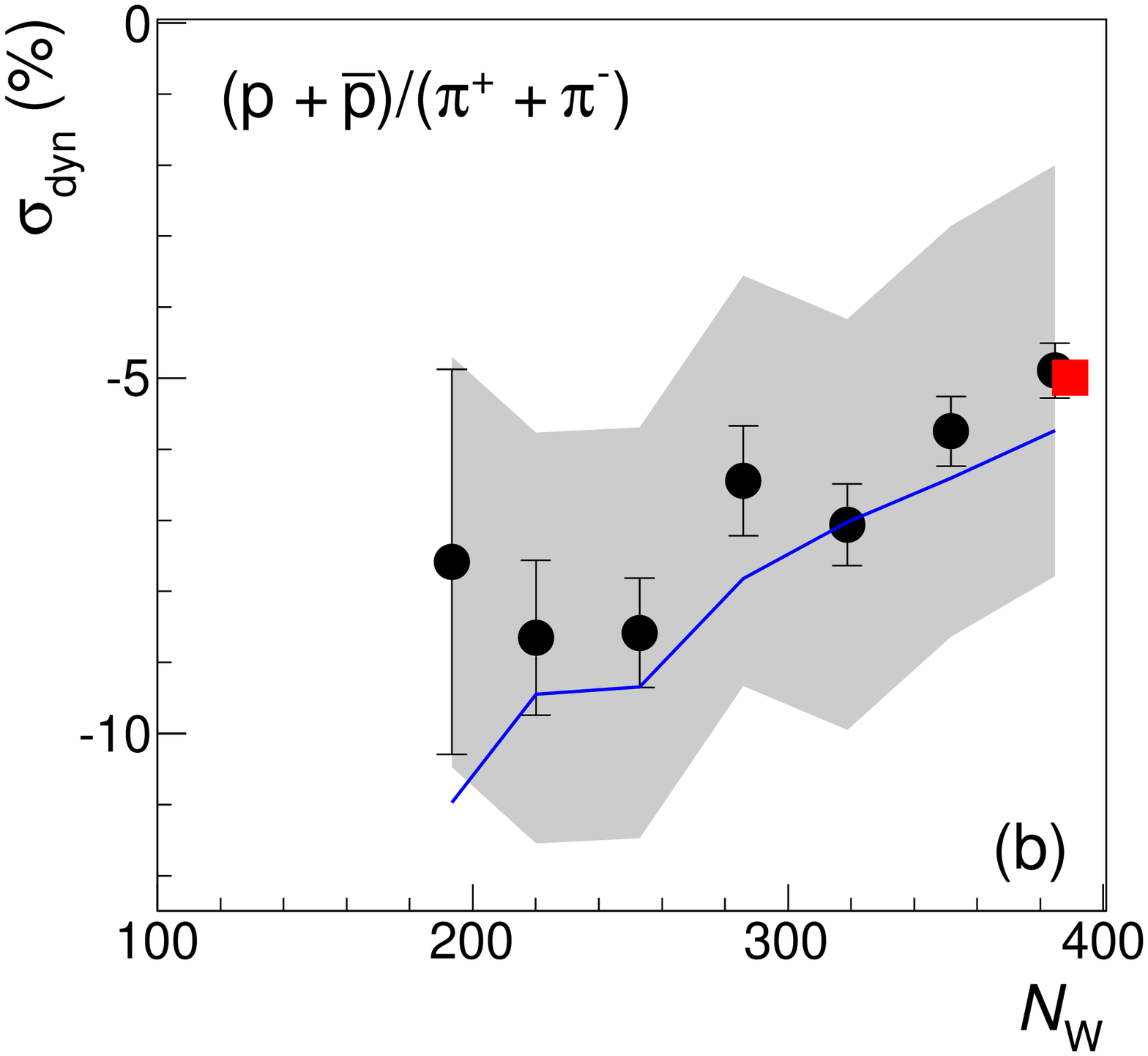}
\includegraphics[width=0.3\textwidth]{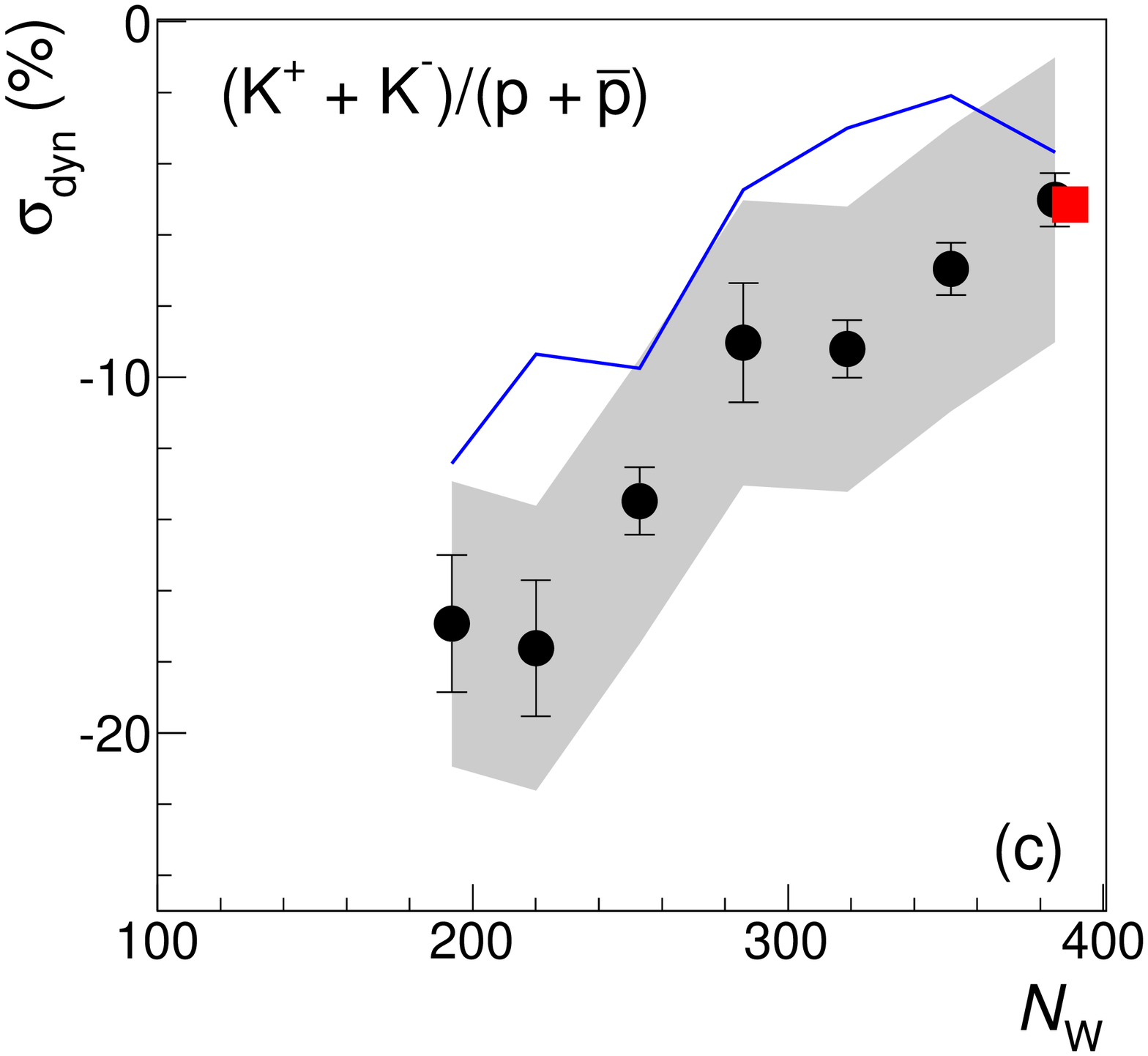}
\caption{(Color online) Centrality dependence of the measure 
 \sdyn~of K/$\pi$ (a), p/$\pi$ (b), and K/p (c) ratio fluctuations. Dots show results
 from this analysis, squares show previously published measurements~\cite{bib:na49:ebefluct,bib:na49:tim}.
 The curves depict predictions of the UrQMD model~\cite{urqmd} for the NA49 acceptance. 
 The shaded bands show the systematic errors.}
\label{fig:diff:all}
\end{figure}

In this section we present our results on the centrality dependence of 
dynamical fluctuations of K/$\pi$, p/$\pi$, and K/p ratios in Pb + Pb collisions
at 158\agev~(numerical values are listed in Table~\ref{tab:results}). The dot symbols
in Fig.~\ref{fig:diff:all} show the dependence of \sdyn~(mean value of the results
for tight and loose track cuts) of the three ratios on the average number of
wounded nucleons \anw. The systematic errors are indicated by the shaded bands. Also
shown by square symbols are the values of dynamical fluctuations in central Pb + Pb collisions
at 158\agev~beam energy from previous NA49 analyses~\cite{bib:na49:ebefluct,bib:na49:tim}, which used a
different event ensemble. The results from both analyses are in good agreement.

One observes the same trend for all considered particle ratios, namely that
the absolute value of the dynamical fluctuations increases
with decreasing centrality (decreasing \nw). Interestingly,
the UrQMD model~\cite{urqmd} reproduces this behavior for all
three ratios as demonstrated by the lines in Fig.~\ref{fig:diff:all}. The model was
previously found to fail in describing the energy dependence of \sdyn$(\text{K}/\pi)$ and
\sdyn$(\text{K}/\text{p})$ whereas it reproduced 
\sdyn$(\text{p}/\pi)$~\cite{bib:na49:ebefluct,bib:na49:tim}.

\begin{table}[htb]
\caption[]{\label{tab:results}Numerical results for \sdyn$(\text{K}/\pi)$,
\sdyn$(\text{p}/\pi)$, and \sdyn$(\text{K}/\text{p})$ with statistical
and sytematic uncertainties for seven centrality intervals in Pb+Pb collisions
at 158\agev. Also listed are the corresponding average number of wounded nucleons $\langle \text{N}_\text{W} \rangle$
and the average numbers of identified particles $\langle \pi \rangle$, $\langle \text{K} \rangle$,
$\langle \text{p} \rangle$ in the acceptance used for the analysis.
}
\begin{center}
\begin{tabular}{c|ccc|ccc}
\hline
\hline
 \anw  & $\sigma_{\text{dyn}}(\text{K}/\pi)$ & $\sigma_{\text{dyn}}(\text{p}/\pi)$ & $\sigma_{\text{dyn}}(\text{K}/\text{p})$ & $\langle \pi^+ + \pi^- \rangle$ & $\langle \text{K}^+ + \text{K}^- \rangle$ & $\langle \text{p} + \bar{\text{p}} \rangle$\\
\hline
384 & $ 3.7 \pm 0.8 \pm 1.2$ & $-4.9 \pm 0.4 \pm 2.9$ & $ -5.0 \pm 0.7 \pm 4.0$ & 349.4 & 45.4 & 51.1 \\
352 & $ 4.0 \pm 1.1 \pm 1.2$ & $-5.7 \pm 0.5 \pm 2.9$ & $ -7.0 \pm 0.7 \pm 4.0$ & 284.5 & 35.3 & 39.3 \\
319 & $ 6.1 \pm 0.9 \pm 1.2$ & $-7.1 \pm 0.6 \pm 2.9$ & $ -9.2 \pm 0.8 \pm 4.0$ & 234.0 & 27.7 & 31.1 \\
286 & $ 3.0 \pm 1.7 \pm 1.2$ & $-6.4 \pm 0.8 \pm 2.9$ & $ -9.0 \pm 1.7 \pm 4.0$ & 191.7 & 21.6 & 23.1 \\
253 & $ 7.0 \pm 1.5 \pm 1.2$ & $-8.6 \pm 0.8 \pm 2.9$ & $-13.5 \pm 0.9 \pm 4.0$ & 156.4 & 16.5 & 18.1 \\
220 & $ 6.9 \pm 1.9 \pm 1.2$ & $-8.7 \pm 1.1 \pm 2.9$ & $-17.6 \pm 1.9 \pm 4.0$ & 124.1 & 12.0 & 13.7 \\
193 & $11.5 \pm 1.5 \pm 1.2$ & $-7.6 \pm 2.7 \pm 2.9$ & $-16.9 \pm 1.9 \pm 4.0$ &  97.7 &  8.5 & 11.3 \\
\hline
\hline
\end{tabular}
\end{center}
\end{table}

\subsection{Scaling behaviour of dynamical fluctuations}

In this section we discuss various multiplicity scaling prescriptions
which were proposed~\cite{bib:jeon,Abelev:kpifluct,Koch:2009dg} with the aim
of separating effects of changing average particle multiplicities
from the energy and collision volume dependence of genuine dynamical fluctuations.
It is important to note that for comparisons of experimental data
with scaling calculations the measured multiplicities inside the 
experimental acceptances should be used. The analysis
will be applied simultaneously to the centrality dependence reported in this
paper and the energy dependence previously published in~\cite{bib:na49:ebefluct,bib:na49:tim}. 

\begin{figure}[ht]
\centering
\includegraphics[width=0.30\textwidth]{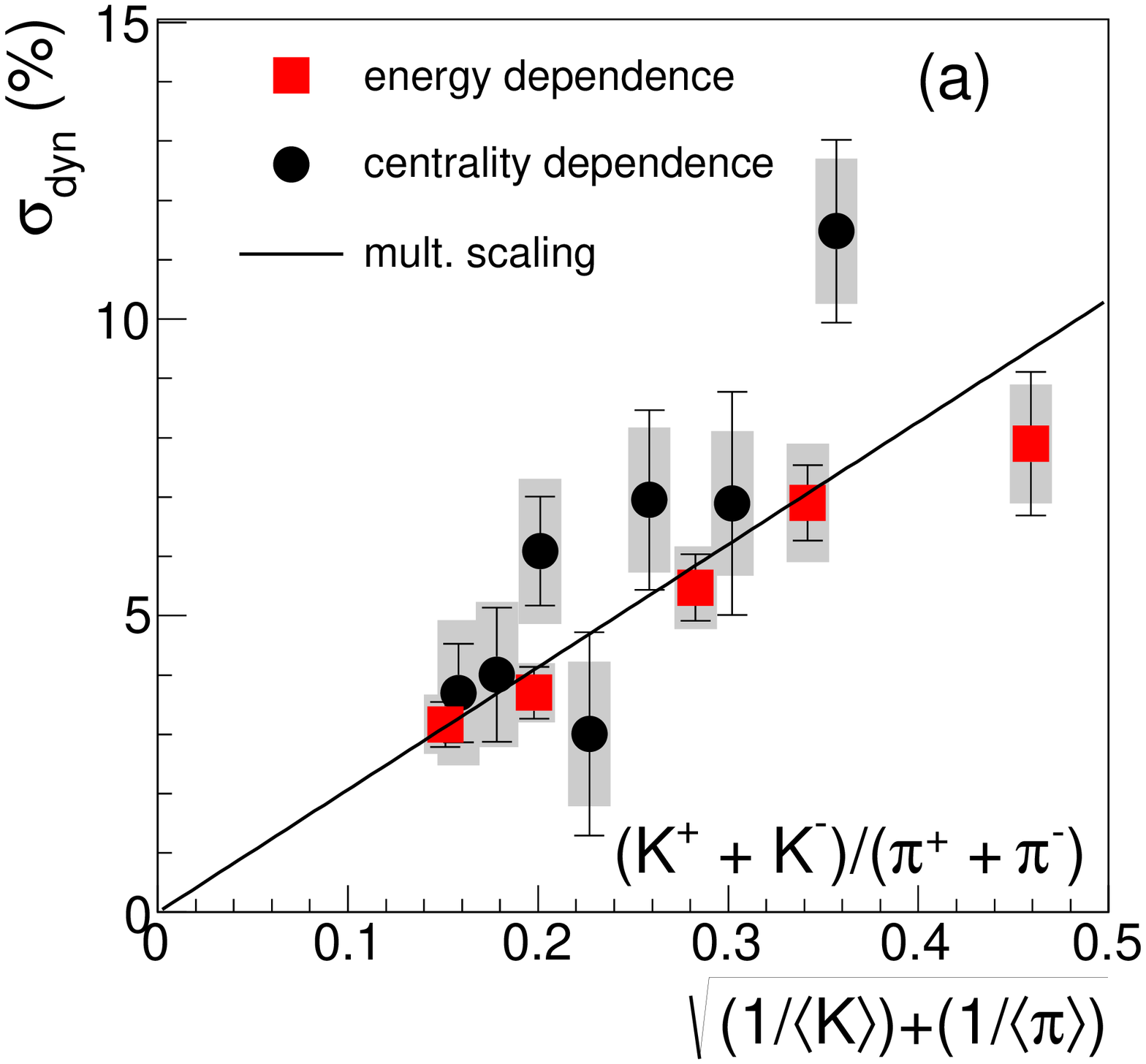}
\includegraphics[width=0.30\textwidth]{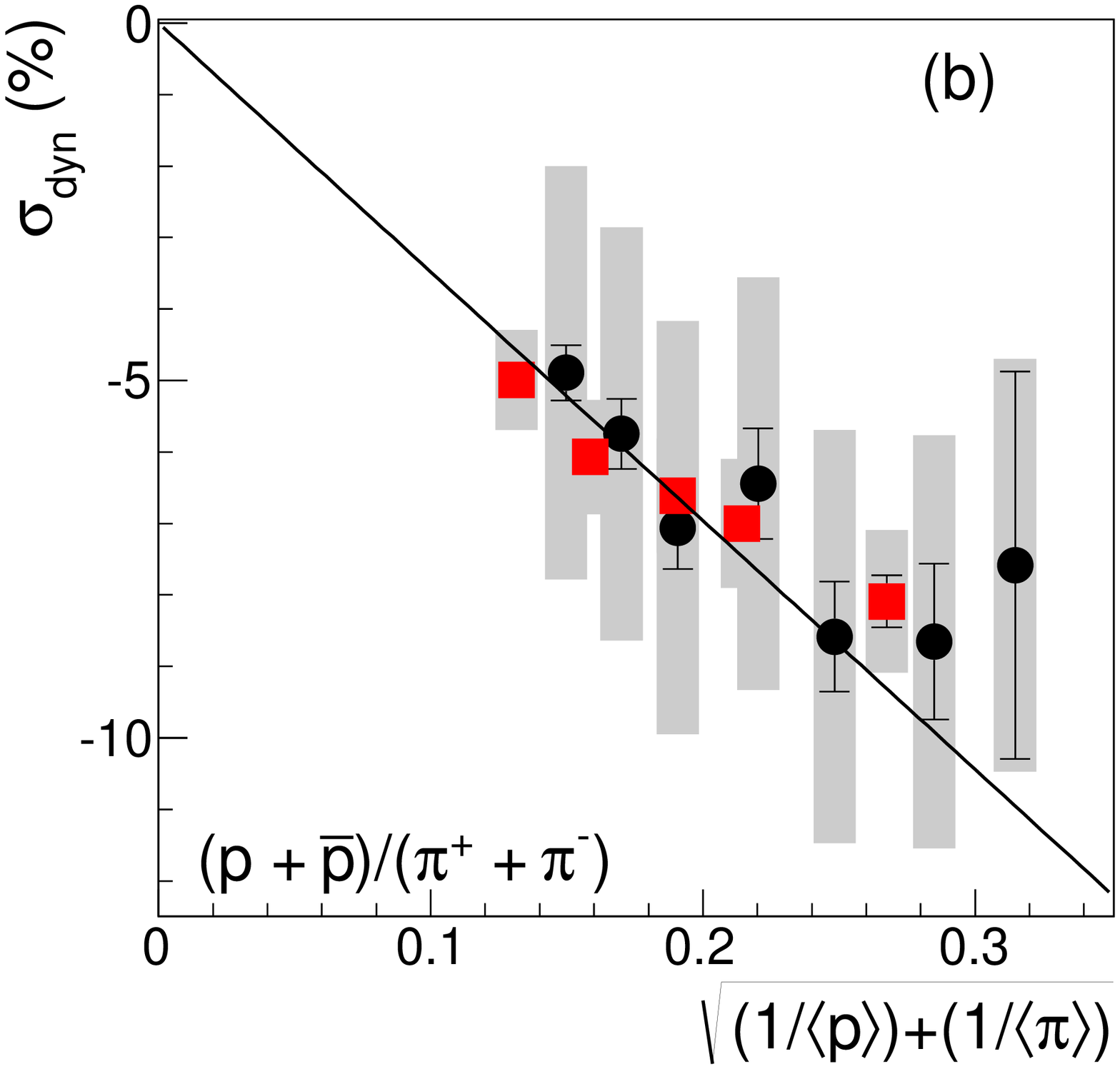}
\includegraphics[width=0.30\textwidth]{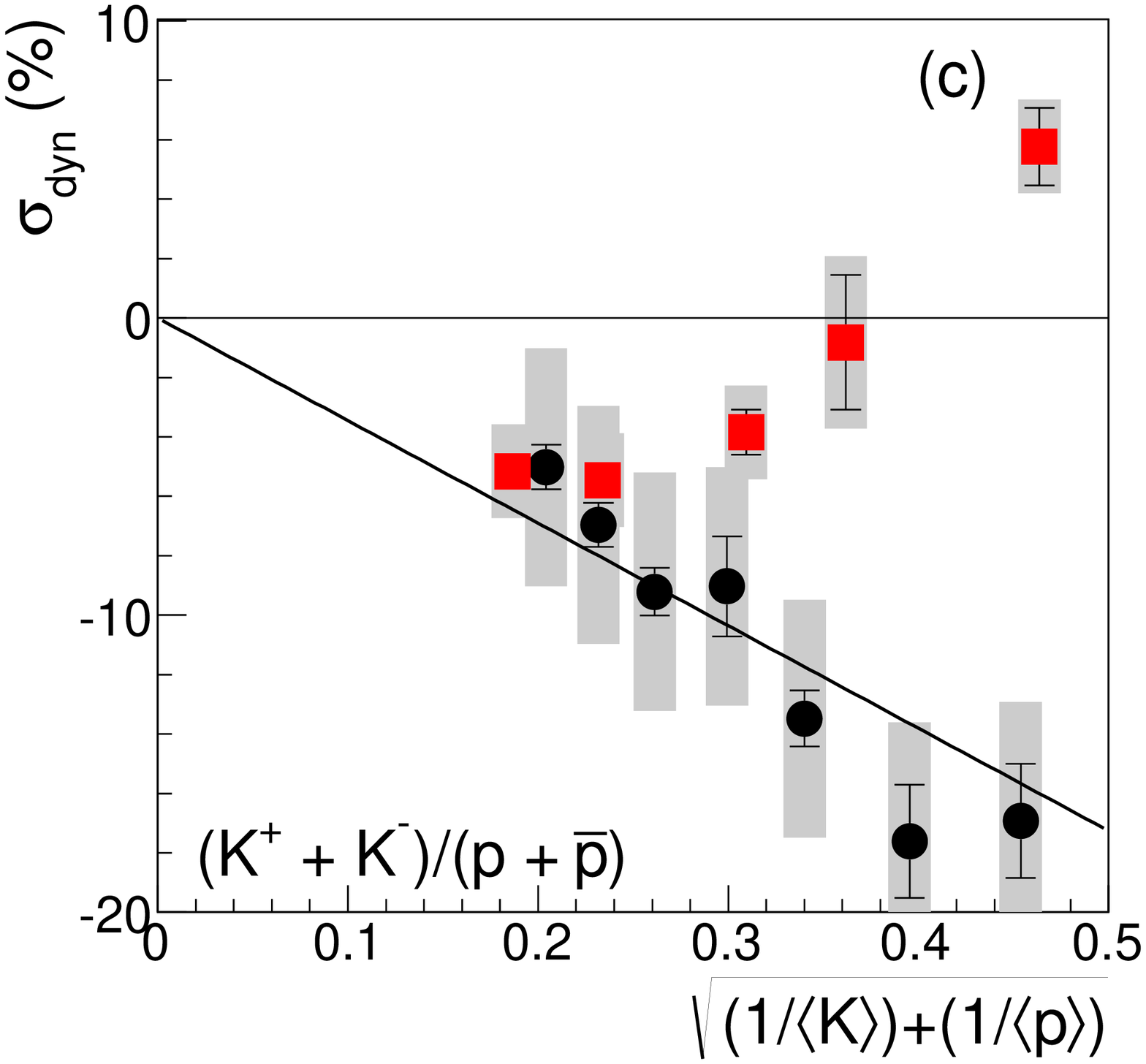}
\caption{(Color online) Dynamical fluctuations of the 
 K/$\pi$ (a), p/$\pi$ (b), and K/p (c) ratio as a function of
 $\sqrt{1/\langle \text{K} \rangle + 1/\langle \pi \rangle}$, 
 $\sqrt{1/\langle \text{p} \rangle + 1/\langle \pi \rangle}$, and 
 $\sqrt{1/\langle \text{K} \rangle + 1/\langle \text{p} \rangle}$ respectively.
 $\langle \pi \rangle$, $\langle \text{K} \rangle$, and $\langle \text{p} \rangle$ are the 
 average number of kaons and protons in the acceptance.
 The solid lines show fits to Poisson multiplicity scaling 
 $\sigma_{\text{dyn}} \propto \sqrt{1/\langle A \rangle + 1/\langle B \rangle}$ (see text).
 Shaded bands indicate systematic uncertainties.}
\label{fig:ktopr:ptopi:ktop}
\end{figure}

In~\cite{Koch:2009dg} it was shown that \sdyn~is expected to have a strong multiplicity
dependence and might scale with $\sqrt{1/\langle A \rangle + 1/\langle B \rangle}$, 
where $\langle A \rangle$ and $\langle B \rangle$ are the average numbers
of accepted particles of type A and B. As shown in Fig.~\ref{fig:ktopr:ptopi:ktop}(a,b)
the measurements of the centrality and energy dependence of K/$\pi$ and p/$\pi$
fluctuations are consistent with the proposed scaling (sometimes called Poisson scaling).
This result suggests that a large contribution to the observed variations appears to
be caused by the changing multiplicities rather than by changes of the underlying
correlations.

In contrast, the energy and centrality dependence of K/p ratio fluctuations, 
plotted in Fig.~\ref{fig:ktopr:ptopi:ktop}(c) as a function of 
$\sqrt{1/\langle \text{K} \rangle + 1/\langle \text{p} \rangle}$,
are not compatible with a common multiplicity scaling.
The energy dependence shows a change
of sign, indicating a change in the underlying correlation around
30\agev~beam energy. On the other hand, the centrality
dependence exhibits a smooth decrease which is
close to the Poisson multiplicity scaling behaviour
(solid line in Fig.~\ref{fig:ktopr:ptopi:ktop}(c)). 

As already mentioned in the introduction section the STAR collaboration
has presented results on the collision energy dependence of particle-ratio fluctuations measured at RHIC
in Au+Au collisions in terms of the observable \ndyn. First results
for $\sqrt{s_{NN}}=7.7$~GeV~\cite{Star:qm2011} presented at conferences 
show a trend which differs from our results for K/$\pi$ and K/p when compared using the relation
between \ndyn~and \sdyn~(see Eq.~\ref{eq:sdyn:nudyn} in section 3.4).
Intensive disussion could not yet determine the cause
of the difference. However, we note that the acceptance in both rapidity $y$ and 
transverse momentum \pt~as well as selection of collision centrality are not the same.

Another scaling behaviour of the dynamical fluctuations of the p/$\pi$ ratio
was proposed in \cite{bib:dkresan} based on the hypothesis that these
originate from the production and decay of nucleon resonances.
Such decays introduce correlations between p and $\pi$. Assuming that the
variance terms in Eq.~\ref{eq:sdyn:nudyn} can be neglected, the
corresponding \sdyn~can be approximated by the following equation:
\begin{equation}
\sigma_{\text{dyn}} \approx -\sqrt{\frac{\text{Cov}(A,B)}{\langle A \rangle \langle B \rangle}} 
 \propto -\sqrt{\frac{( \langle A \rangle \langle B \rangle)^{\alpha}}
        { \langle A \rangle \langle B \rangle}} \text{~~,}
\label{eq:prot2pi}
\end{equation}
with the parameter $\alpha$ expected to have the value 0.5.

\begin{figure}[ht]
\centering
\includegraphics[width=0.35\textwidth]{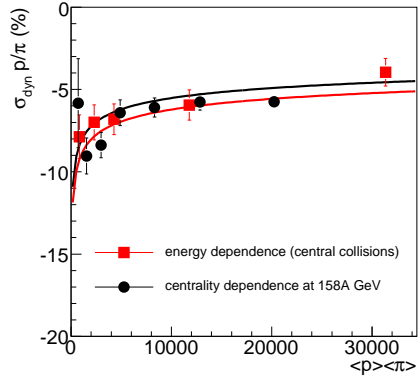}
\caption{(Color online) Dynamical fluctuations of the p/$\pi$ ratio as a function of
the product of average numbers of protons and pions in the detector
acceptance.}
\label{fig:scaling:p2pi}
\end{figure}

The energy and centrality dependences of the dynamical fluctuations of the
p/$\pi$ ratio expressed as functions of the product $\langle\text{p}\rangle\langle\pi\rangle$ are plotted
in Fig.~\ref{fig:scaling:p2pi}. 
A fit of the data points with Eq.~\ref{eq:prot2pi}
resulted in $\alpha$ parameters equal to
$\alpha=0.66\pm0.12$ for the energy dependence and $\alpha=0.51\pm0.03$ for the
centrality dependence. This experimental observation supports the
hypothesis that the source of the p/$\pi$ ratio fluctuations is nucleon
resonance production and decay. 

\begin{figure}[ht]
\centering
\includegraphics[width=0.35\textwidth]{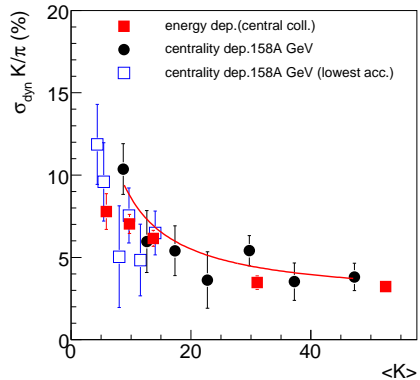}
\caption{(Color online) Dynamical fluctuations of the K/$\pi$ ratio as a function of
average number of kaons in the detector acceptance. Open square symbols show the
results for the acceptance of the most peripheral set of events.
The solid line shows the fit of the centrality dependence with the function
of Eq.~\ref{fig:ka2pi:fit}.}
\label{fig:ka2pi:nka}
\end{figure}

An alternative scaling hypothesis was also investigated for the K/$\pi$ ratio fluctuations. 
Since $\langle \text{K} \rangle \ll \langle \pi \rangle$,
the dominating term in Eq.~\ref{eq:sdyn:nudyn} for the dynamical fluctuations of the
K/$\pi$ ratio may be the kaon variance term, 
provided the covariance term can be neglected.
Fig.~\ref{fig:ka2pi:nka} shows the energy and centrality dependence
of the K/$\pi$ ratio fluctuations versus the number of kaons $\langle$K$\rangle$ in the acceptance. 
The curves in Fig.~\ref{fig:ka2pi:nka} indicate that also the function
\begin{equation}
f(\langle \text{K} \rangle)  = a + \frac{b}{\langle \text{K} \rangle}
\label{fig:ka2pi:fit}
\end{equation}
provides a good fit to both the centrality and energy dependence of K/$\pi$ ratio fluctuations
with $a = 2.4 \pm 0.8$ and $b = 62.1 \pm 16.6$.

In the most peripheral collision events fewer phase space bins are useable
because of the lower multiplicities. We repeated the analysis by
restricting the extraction of \sdyn~to this smaller
acceptance for all centralities. The multiplicities $\langle$K$\rangle$
in the restricted acceptance, of course, decrease. Nevertheless,
the open square symbols in Fig.~\ref{fig:ka2pi:nka} demonstrate that the results
for \sdyn~still follow the scaling of Eq.~\ref{fig:ka2pi:fit}.

Presently the NA49 collaboration is in the process of developing and applying a new 
analysis procedure (identity method~\cite{identity}) for the determination of 
event-by-event particle ratio fluctuations. It is designed to unfold the
second moments of the multiplicity distributions of protons, kaons and
pions. With this information, more direct tests of
various models will become possible.


\section{Summary}

We presented new measurements of the centrality dependence of p/$\pi$, K/$\pi$, and K/p particle 
ratio fluctuations in terms of \sdyn~obtained by the NA49 experiment from Pb + Pb collisions 
at 158\agev. The measure \sdyn~increases
in absolute value with decreasing centrality for all these ratios. Comparisons to
various multiplicity scaling schemes were made to both the 
centrality and the previously published energy dependences.
Fluctuations of the p/$\pi$ and K/$\pi$ ratios are consistent with Poisson multiplicity scaling,
thus suggesting that changing multiplicities rather than varying genuine correlations
are the main source of these dependences.
The p/$\pi$ ratio fluctuations also scale with
1/($\langle\text{p}\rangle\langle\pi\rangle$)$^{0.5}$, supporting the assumption that they originate from production
and decay of nucleon resonances. The
K/$\pi$ ratio fluctuations are also compatible with a 1/$\langle\text{K}\rangle$ behavior, suggesting
fluctuations of the kaon multiplicity as the main source of the measured energy and centrality dependences.
In contrast, multiplicity scaling cannot describe the measurents of K/p fluctuations consistently.
Although the centrality dependence of the absolute value of K/p ratio fluctuations
exhibits a smooth increase for more peripheral collisions and is compatible with
Poisson multiplicity scaling, a sign
change is observed for the energy dependence. Therefore, the correlations
causing K/p fluctuations appear to be changing in the SPS energy range.

\section{Acknowledgements}
Acknowledgements: This work was supported by
the US Department of Energy Grant DE-FG03-97ER41020/A000,
the Bundesministerium fur Bildung und Forschung, Germany (06F~137),
the Virtual Institute VI-146 of Helmholtz Gemeinschaft, Germany,
the Polish Ministry of Science and Higher Education (1~P03B~006~30, 1~P03B~127~30, 
0297/B/H03/2007/33, N~N202~078735,  N~N202~078738, N~N202~204638),
the Hungarian Scientific Research Foundation (T032648, T032293, T043514),
the Hungarian National Science Foundation, OTKA, (F034707),
the Bulgarian National Science Fund (Ph-09/05),
the Croatian Ministry of Science, Education and Sport (Project 098-0982887-2878)
and
Stichting FOM, the Netherlands.



\end{document}